\documentclass[12pt]{article}

\usepackage{sbc-template}

\usepackage{cite}
\usepackage{amsmath,amssymb,amsfonts}
\usepackage{algorithmic}
\usepackage{graphicx}
\usepackage{textcomp}
\usepackage{xcolor}
\usepackage{colortbl}
\usepackage{lipsum}
\usepackage{subcaption}
\usepackage{pgfplots}
\usepackage{fancyhdr}
\usepackage{lastpage}
\usepackage{makecell}
\usepackage{dblfloatfix}
\usepackage{soul}
\usepackage{booktabs}
\usepackage{comment}
\usepackage{scalefnt}
\usepackage{multirow}
\usepackage{bigstrut}
\usepackage{scalefnt}
\usepackage{hyperref}

\pgfplotsset{compat=1.8}

\def\BibTeX{{\rm B\kern-.05em{\sc i\kern-.025em b}\kern-.08em
    T\kern-.1667em\lower.7ex\hbox{E}\kern-.125emX}}

\title{Evaluating the Performance of Twitter-based Exploit Detectors}

\author{Daniel Alves de Sousa\inst{1}, Elaine Ribeiro de Faria\inst{1}, Rodrigo Sanches Miani\inst{1} }

\address{School of Computer Science (FACOM) -- Federal University of Uberlândia (UFU)\\
Uberlândia, MG, Brazil
  \email{\{danielsousa,elaine,miani\}@ufu.br}
}

\begin{document}

\maketitle

\definecolor{Gray}{gray}{0.9}
\newcolumntype{g}{>{\columncolor{Gray}}r}
\newcolumntype{G}{>{\columncolor{Gray}}c}

\begin{abstract}
Patch prioritization is a crucial aspect of information systems security, and knowledge of which vulnerabilities were exploited in the wild is a powerful tool to help systems administrators accomplish this task. The analysis of social media for this specific application can enhance the results and bring more agility by collecting data from online discussions and applying machine learning techniques to detect real-world exploits. In this paper, we use a technique that combines Twitter data with public database information to classify vulnerabilities as exploited or not-exploited. We analyze the behavior of different classifying algorithms, investigate the influence of different antivirus data as ground truth, and experiment with various time window sizes. Our findings suggest that using a Light Gradient Boosting Machine (LightGBM) can benefit the results, and for most cases, the statistics related to a tweet and the users who tweeted are more meaningful than the text tweeted. We also demonstrate the importance of using ground-truth data from security companies not mentioned in previous works.
\end{abstract}
     

\section{Introduction}
\label{sec:introducao}

Exploit detection is an essential application for system administrators as system updates sometimes involve rigorous impact analysis and even severe adaptations or migrations. An exploited vulnerability on a particular system should demand urgent responses from its administrator in the sense of prioritizing patches. Given this scenario and the fact that many vulnerabilities might never be exploited in real-world attacks \cite{someVulnerabilitiesAre}, knowledge of which vulnerabilities are more likely to be exploited in the wild can be an excellent tool for system administrators to prioritize patch deployments. There are a few metrics that could be used with this purpose (such as the Common Vulnerability Score System - CVSS base scores and the Microsoft Update Severity Rating System), but they err on the side of caution \cite{comparing-cvss-ms}. The analysis of social media data can leverage this process by taking advantage of the community's discussions on the topic \cite{shrestha2020multiple}. 


    The work presented in \cite{Suciu:2015} has shown that hackers, system administrators, and software vendors discuss vulnerabilities on social media like Twitter. The authors also showed, for the first time, that this information could be used to create a framework for predicting exploits using machine learning techniques. Several works published after that investigated the feasibility of such \textit{Twitter-based early exploit detectors} (\cite{Bullough2017}, \cite{Queiroz2017}, and \cite{predict-when}). Despite these works shown the potential of using Twitter data to detect exploits, they still have limitations. References \cite{Bullough2017} and \cite{Queiroz2017}, for example, uses only Support Vector Machine (SVM) as its classifier algorithm. \cite{predict-when} partially solves this issue by adding several other classifiers in their study. However, they do not provide a performance comparison using the original dataset described in \cite{Suciu:2015}. Another issue is related to the use of a single source, Symantec Intrusion Protection Signature, for building the ground-truth of real-world exploits. As discussed in \cite{Suciu:2015}, this is a notable limitation since Symantec does not cover all platforms and products uniformly.
    
      
   
Moreover, none of the previous work evaluates the impact of training Twitter-based exploit detectors using past data to predict the future. For example, suppose that an exploit detector was trained using data from 2017. What happens if only data from 2018 would be presented to this model? Would add more training data gives better performance? This discussion is important to evaluate the practical implications of using Twitter data to build exploit detectors and consequently helping prioritize which vulnerabilities to patch. Based on these analyses, we propose to evaluate the extent to which different factors influence the performance of Twitter-based exploit detectors. We focus on exploring some machine learning characteristics, training classifiers with different time-window sizes, and evaluating the impact of ground-truth labels from different sources. 
  
The paper has four main contributions. First, we identify a suitable classifier for building Twitter-based exploit detectors using a five-year dataset composed of tweets and vulnerability information. Second, we develop a ground-truth for labeling real-world exploits using data from sources other than Symantec. Third, we provide empirical evidence that using ground-truth information from a single vendor can bias the model toward some vulnerabilities and induce to non-optimal performance on real-world scenarios. Fourth, we examine if the performance of an exploit detector model is affected along the time. Our results suggest that models trained and tested using data from a single calendar year outperform those trained with data from previous years. This indicates that selecting the right amount of past information that will feed the model is decisive to improve its performance. 

    The rest of this paper is organized as follows. Section \ref{sec:background} presents the basic terminology about security vulnerabilities. Section \ref{sec:related-work} reviews and compares related studies with this work. Section \ref{sec:proposal-experiments} details the dataset, features and classifiers that are used in our system architecture. Section \ref{sec:results} presents the results and shows some threats to validity. Finally, Section \ref{sec:conclusion-future} concludes the paper and suggests future work.

\section{Terminology} \label{sec:background}

  
Common Vulnerabilities and Exposures (CVE) is a list maintained by MITRE which assigns a unique number to each disclosed vulnerability. Meltdown vulnerability, for instance, is identified by the number CVE-2017-5754. Aside from the official channels, some vulnerabilities may also be disclosed through forums, social media, or blogs, which may lead to a situation where the CVE of a non-patched flaw can be published. In either scenario, official and non-official disclosure, malicious hackers can take advantage of a vulnerability to harm unpatched systems. \emph{Exploit} is the term used to define the techniques or tools developed with this goal.
    
Exploits can be divided into two categories: proof of concept (PoC) and real-world (RW) exploits \cite{Suciu:2015}. While the first is developed as part of the disclosure process to demonstrate a particular vulnerability, the latest is created to perform real attacks. Although some PoC may be used in real-world scenarios, others are too impractical to be. Therefore, vulnerabilities with exploits in the wild are a subset of the ones with PoC exploits.

 

    

\section{Related Work} \label{sec:related-work}

Many previous works have addressed the task of using machine learning to predict whether a vulnerability will be exploited or not. \cite{bozorgi} 
trained an SVM classifier using features extracted from the Open Source Vulnerability Database (OSVDB) and the NVD to predict if a vulnerability is likely to be exploited. As ground truth, the authors used a metric called ``Exploit Classification'' from the OSVDB, no longer available since 2016. Despite getting nearly 90\% of accuracy, the ground truth used presents a very high positive rate (exploited vulnerabilities), contrasting with most related works (\cite{someVulnerabilitiesAre}, and \cite{beforeWeKnewIt}, for example).

\cite{Suciu:2015} introduced the use of Twitter to help classifying vulnerabilities as exploited or not exploited. Like \cite{bozorgi}, the authors acquired data from the NVD and the OSVDB, but they added an extra set of features extracted from Twitter, including text and statistics about tweets and users who tweeted. The work divides the ground truth into two groups, PoC and RW exploits, using the Exploit Database (EDB) \footnote{https://www.exploit-db.com} as a source for PoC and Symantec's antivirus and IPS signatures for RW. They collected tweets from February 2014 and January 2015 and found evidence that the use of Twitter data could increase the classifier's overall performance. The paper, however, does not explore other classifiers options (only SVM was used) or methods to overcome dataset imbalance, it uses only one year's worth of data, and relies on a single antivirus vendor for RW ground truth information. \cite{Queiroz2017} used a similar approach to detect useful information about security vulnerabilities using Twitter data. They collected posts from security specialists from March 2016 to early March 2017 and manually labeled training data. Despite not being the focus of the paper, the approach was able to identify useful alerts about vulnerability exploits. 


\cite{Bullough2017} raised questions about prior work's methodology and highlighted how small changes in using the dataset could affect the performance of predictive models. The authors have been especially critical about temporal intermixing caused by random splitting data for train and test. Such ideas are valuable and should be considered when planning new models, but their conclusion about using a temporal split may not be accurate. In Section \ref{sec:res_time_window}, we reproduce this test in a variety of ways, and our results suggest that performance differences may have different reasons. Furthermore, the work uses only PoC ground truth from EDB and found 18\% of their CVEs exploited, a value significantly above those presented in works about RW exploits.

\cite{predict-when} used an ensemble of regression algorithms to predict when a vulnerability will be exploited, both for PoC and RW scenarios. As features, the authors created a graph-based model relating CVE-Authors-Tweet and, for ground truth, only Symantec's data was used. The authors also approached the temporal intermixing issue, demonstrating how, in some cases, the CVSS is not available at the time of the vulnerability's disclousure. In our work, we will demonstrate in section \ref{sec:res_algs} that the CVSS plays a small part in the classifier's performance. Nonetheless, those issues may indicate that more relevant NVD data could be affected similarly and should also be studied.   

\section{Proposal and Experiments}\label{sec:proposal-experiments}

In this work, we evaluate and propose improvements in the Twitter-based exploit detection method presented in \cite{Suciu:2015}. We chose to use that paper as our baseline because other related works were not entirely comparable to the best of our knowledge, displaying very different rates of exploited vulnerabilities or using completely different data sources. We start by using the same dataset from the mentioned work, which contains messages posted on Twitter, together with public data, and we experiment with different machine learning techniques to classify if a vulnerability will be exploited. We then extend the ground truth with information from different anti-malware software, and finally, we extend the dataset with data from 2015 to 2018. In all cases, we only consider already cataloged vulnerabilities (those to which a CVE number was assigned). To summarize, this paper has four main goals:
\begin{itemize}
    \item Compare classification algorithms: we want to compare the performance of four well-known algorithms on classifying vulnerabilities as ``exploited'' or ``not exploited'' based on data from Twitter and public vulnerabilities databases. Namely, we compared Support Vector Machines, Logistic Regression, XGBoost, and LightGBM. We also intend to evaluate how different groups of features impact each algorithm. In our method, we divide features into four groups: Twitter text, Twitter metadata, CVSS score and subscores, and a set of data from public vulnerabilities databases (mainly from the NVD).
    \item Compare Multiple Ground Truth: previous works rely only on Symantec's antivirus and intrusion protection system (IPS) signatures to indicate real-world exploits. We want to evaluate if other antivirus databases can provide useful insight and improve prediction performance.
    \item Class Balancing: we want to verify if class balancing methods can improve the overall results, given that class imbalance is one of the main challenges of this task.
   \item Updated Data and Different Time Window Sizes: we want to evaluate how the method behaves on more recent data and understand how the classifier is affected by temporal splits and changes in the volume of training and testing data. To do that, we create time windows with different sizes covering different periods to train and test our model.
\end{itemize}

Our classifier considers each CVE as an instance to which should be assigned \emph{true}, if the vulnerability was exploited, or \emph{false} otherwise. The features used to characterize each instance summarize the data collected about a specific CVE. They contain a Bag-of-Words (BoW) representation of tweets mentioning that CVE, Twitter statistics and metadata related to those tweets, and public database information about the vulnerability. In Section \ref{subsec:set-features} we detail these features. 

To train a classifier, we need a way to resolve if a vulnerability has any known exploit. On the first test, we use the same ground truth data as defined in \cite{Suciu:2015}: the ExploitDB (EDB) and Symantec's antivirus and intrusion protection system (IPS) signatures. For all remaining tests, we improve the ground truth with data from Avast\footnote{https://www.avast.com/exploit-protection.php}, ESET\footnote{https://www.virusradar.com/en/threat\_encyclopaedia}, and Trend Micro\footnote{https://www.trendmicro.com/vinfo/us/threat-encyclopedia}. The EDB is an online resource of known exploits that also provides information about the vulnerabilities affected. While the EDB is an excellent resource of PoC exploits, an antivirus signature is probably the best indicator that an exploit has been spotted in the wild. Past works only included Symantec's database as a source for such information, but we were able to find similar data from other vendors. In Section \ref{sec:res_gts}, we analyze the quality of data and how they can impact past results.

\subsection{System's Architecture}

\begin{figure}[ht]
    \centering
    \includegraphics[width=.52\textwidth]{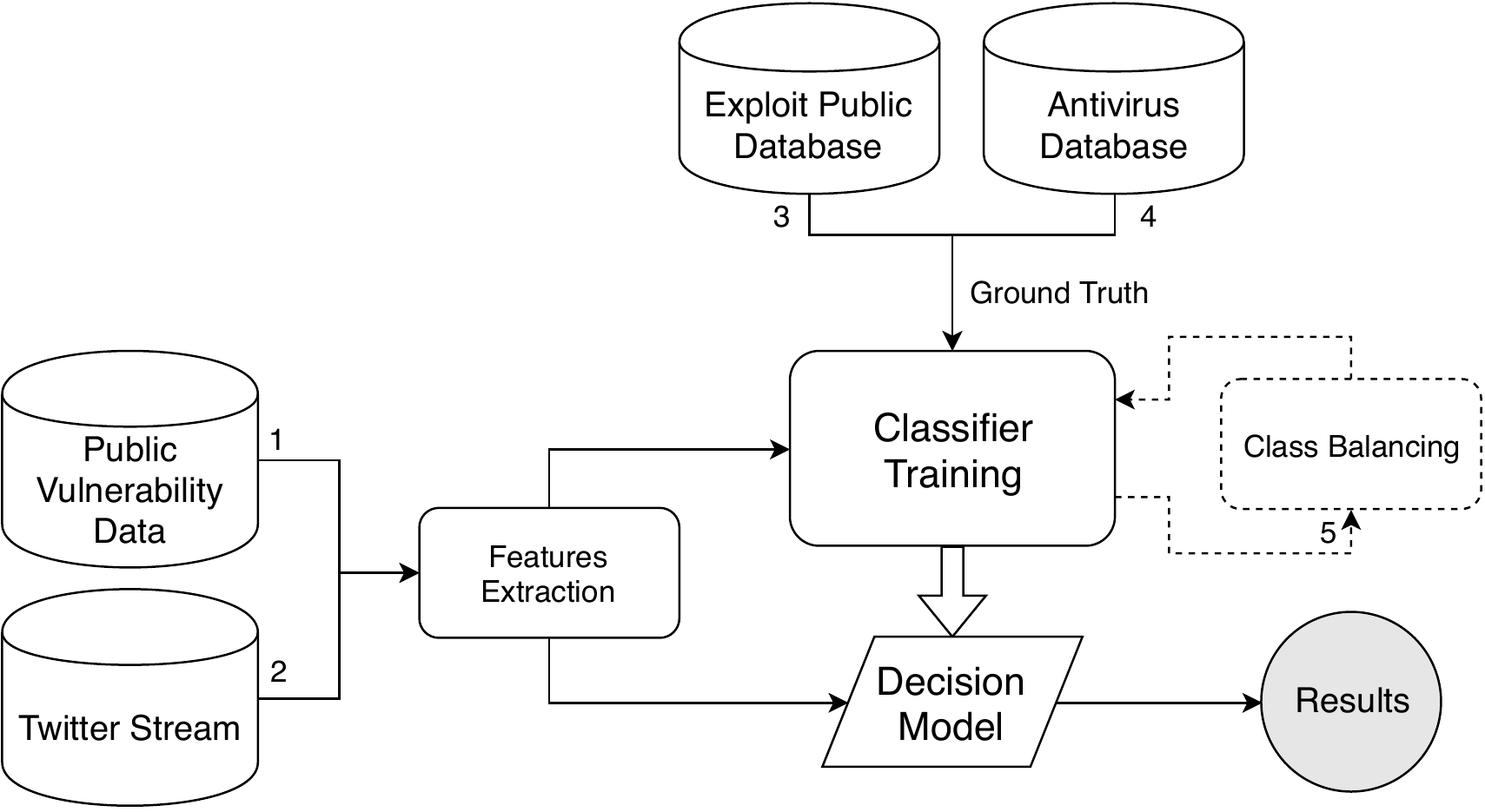}
    \caption{System Architecture}
    \label{fig:system-arch}
\end{figure}

    Fig. \ref{fig:system-arch} represents the system used in this work. Data gathering process and its sources (numbers 1 to 4) are detailed in subsection \ref{subsec:dataset}. Feature extraction process is presented in \ref{subsec:set-features} while the machine learning techniques and the balancing phase (number 5) are clarified in \ref{subsec:classifier}. We developed several Python scripts for supporting the data gathering and feature extraction process. We also use the \emph{scikit-learn} library \cite{scikit-learn} for conducting the classification tasks.

\begin{table}[h]
\scalefont{.7}
    \caption{Vulnerabilities mentioned on Twitter}
    \label{tab:cves-numbers}
    \centering
    \begin{tabular}{|r|c|c|c|}
    \hline
         & \# of Vulnerabilities & Mentioned on Twitter & \%   \\ \hline
    2015 & 6484                  & 822                  & 13\% \\ \hline
    2016 & 6447                  & 776                  & 12\% \\ \hline
    2017 & 14714                 & 1292                 & 9\%  \\ \hline
    2018 & 16556                 & 3753                 & 23\% \\ \hline
    \end{tabular}
\end{table}

\subsection{Dataset}\label{subsec:dataset}

    For the first part of this work, we used the same dataset described in \cite{Suciu:2015} as it was a good baseline for comparison. In essence, it consists of tweets collected using Twitter's Streaming API from February 2014 to January 2015 using the keyword ``CVE''. Each tweet was automatically associated with a vulnerability using the CVE number, and then, to each vulnerability, detailed data (such as vendor, CVSS, and description) were collected from the NVD and the Open Source Vulnerability Database (OSVDB). The authors were able to collect 287,717 tweets containing explicit references to CVE numbers and referencing 5,865 different CVEs from the period studied. 
    
    
    For the second part of our research, we collected tweets containing the word ``CVE'' from January 2015 to December 2018, filtered out those referring to vulnerabilities outside our test period and those not mentioning a valid CVE number. We were able to find 44,570 messages, mentioning 6,643 vulnerabilities discussed by 4,033 users. Table \ref{tab:cves-numbers} shows the number of CVEs mentioned on Twitter by year. In 2018, for example, 23\% of all CVEs disclosed in that year were mentioned at least one time on Twitter.
    
    Unlike \cite{Suciu:2015}, which collected tweets through Twitter Stream API, we collected messages by searching old tweets using the GetOldTweets3 tool \cite{getoldtweets}. We believe this may be the cause of the difference in information volume since accounts and messages deleted will not be found in our method. We also collected data from the NVD for feature extraction. Section \ref{subsec:set-features} contains the list of features to each CVE. As discussed in \cite{Suciu:2015}, we also used two different ground truth for the dataset: one for proof-of-concept (PoC) and another for real-world (RW) exploits. Therefore some of the experiments are also divided into two categories, one for each type of exploit. For PoC we used data from ExploitDB. For the real-worlds exploits, most related studies use signatures from Symantec's antivirus and IPS. We included information from four other vendors: Avast, ESET, Trend Micro, and Kaspersky. We believe that relying on a single vendor can lead to biased results and less efficient learning from the model. In all cases, not all signatures mention the CVE exploited, imposing some limitations on the results.
    


\subsection{Features}\label{subsec:set-features}

For all of our tests, we divided the features used by the classifier into four categories: Twitter text, Twitter statistics and metadata, CVSS score, and Public Vulnerabilities Databases. Twitter text represents a combination of all messages tweeted about a CVE. Table \ref{tab:features} shows a summary of our features. We represent the text with the Bag of Word (BoW) model, and the words chosen were the same as in \cite{Suciu:2015}. The keyword dataset comprises 36 words and some examples include: \textit{0day}, \textit{advisory}, \textit{beware}, \textit{ssl}, and \textit{fix}. Twitter statistics contain data such as the number of retweets related to a CVE and the combined number of followers from all users who tweeted a CVE. Public Vulnerabilities Database originally included information from the NVD and the OSVDB, but we used only the former since the latter is no longer available. The CVSS category contains features representing the CVSS vectors. For the 2015 to 2018 data, we included features for the CVSS 3.0. We also included impact and exploitability subscores for both versions 2.0 and 3.0.

\begin{table*}[h!t]
\scalefont{.7}
    \caption{Summarized List of Features}
    \label{tab:features}
    \begin{center}
        \begin{tabular}{|c|l|l|}
        \hline
        \textbf{Position} & \textbf{Category} & \textbf{Data type} \\\hline
        
		0 - 35	&	Twitter text                        & BoW	    \\\hline
		36 - 47	&	Twitter metadata	                & All Numeric           \\\hline
		48 - 56	&	CVSS Score 2.0                      & 3 Numeric, 6 Categorical Ordinal          \\\hline
		57 - 67	&	CVSS Score 3.0                      & 3 Numeric, 8 Categorical Ordinal          \\\hline
		68 - 78	&	Public Vulnerabilities Databases    &  6 Numeric, 5 Binary		\\\hline
        
        \end{tabular}
    \end{center}
\end{table*}

\begin{table*}[tp!]
\scalefont{.7}
    \centering
        \begin{tabular}{lrrrrr}
        \hline
        \multicolumn{1}{|l|}{}                        & \multicolumn{1}{c|}{\textbf{2014*}} & \multicolumn{1}{c|}{\textbf{2015}} & \multicolumn{1}{c|}{\textbf{2016}} & \multicolumn{1}{c|}{\textbf{2017}} & \multicolumn{1}{c|}{\textbf{2018}} \\ \hline
        \multicolumn{1}{|l|}{\textbf{Mentioned CVEs}} & \multicolumn{1}{r|}{5,865}          & \multicolumn{1}{r|}{822}           & \multicolumn{1}{r|}{776}           & \multicolumn{1}{r|}{1292}          & \multicolumn{1}{r|}{3753}          \\ \hline
        \multicolumn{1}{|l|}{\textbf{Exploited}}      & \multicolumn{1}{r|}{77 (1.3\%)}     & \multicolumn{1}{r|}{71 (8.6\%)}    & \multicolumn{1}{r|}{31 (4.0\%)}    & \multicolumn{1}{r|}{61 (4.7\%)}    & \multicolumn{1}{r|}{133 (4.2\%)}   \\ \hline
        \multicolumn{1}{|l|}{\textbf{PoC}}            & \multicolumn{1}{r|}{383 (6.5\%)}    & \multicolumn{1}{r|}{115 (14.0\%)}  & \multicolumn{1}{r|}{90 (11.6\%)}   & \multicolumn{1}{r|}{220 (17.0\%)}  & \multicolumn{1}{r|}{257 (11.9\%)}  \\ \hline
        \multicolumn{6}{l}{*Sabottke dataset}                                                                                                                                                                                                  
        \end{tabular}
    \textbf{\caption{\label{table:qntd-cves}Number of CVEs exploited compared to total}}
\end{table*}

\subsection{Classifiers}\label{subsec:classifier}

We used the SVM classifier as our baseline since it is well suitable for text categorization \cite{svm-1998-joachims} and because it was used in most related works. We tested several supervised classification algorithms, but we will approach the ones which stood out best: Logistic Regression, XGBoost, and Light Gradient Boosting Machine (LightGBM). We also tried several class balancing algorithms available on Python's \textit{imbalanced-learn API} \cite{imblearn}. In this paper, we will cover only the ones which performed best with our application: Synthetic Minority Over-sampling Technique (SMOTE),  Adaptive Synthetic (ADASYN), Nearest-Neighbor (AllKNN), and the Random Under Sampler (RUS). The first two being over-sampling techniques and the other two under-sampling algorithms.

We used the stratified 10-fold cross-validation and averaged the results. When testing the balancing algorithms, we applied the methods on the training set of each fold. Because our dataset contains features with different data types, and the experimented algorithms also use different types as input, we tested multiple scaling and feature representation methods for each algorithm and used the one which performed best. All categorical-ordinal data (mostly related to the CVSS score) or binary features were first converted to numeric values. Standardization was done through sklearn's \emph{StandardScaler}.

\section{Experimental Results}\label{sec:results}


To compare our results with the original work (\cite{Suciu:2015}), we use as our baseline an SVM classifier with no class balancing and Symantec as the single source of information about real-world exploits. Results are shown using the values of precision ($ \frac{TP}{TP+FP} $, the fraction of correct positive prediction), recall ($ \frac{TP}{TP+FN} $, the fraction of positive cases that were correctly predicted), and F-score ($ 2 * \frac{precision*recall}{precision+recall} $, the harmonic mean of precision and recall). We use the precision-recall (PR) curve as the visual representation for all experiments, considering the database is highly imbalanced. Table \ref{table:qntd-cves} shows how imbalanced the classes are. Since we already have conservative exploits indicators on the CVSS scores, we prioritize the increase of the precision over the recall without sacrificing the F-score. For each algorithm, we tested two scenarios: PoC and RW exploits.


To extract more significant conclusions, we plotted the classifier results when training and testing with each subset of features. We also plot a line for the results using all features combined. By adopting this strategy, it is possible to see how different features may favor a particular algorithm. Furthermore, we ran tests with combinations of subsets, e.g., CVSS and Twitter Statistics, and eventually concluded that certain features are not suitable for some of the algorithms. We have shortened the names of the categories: \textit{Words} stands for Twitter text, \textit{Twitter Stats} stands for Twitter statistics and metadata, \textit{CVSS} stands for the CVSS score and, \textit{Database} stands for Public Vulnerabilities Databases. A description of each subset can be found in Section \ref{subsec:set-features}.

\subsection{Analyzing the Performance of Different Algorithms and Groups of Features}\label{sec:res_algs}
Table \ref{table:alg-results} shows the overall performance of the tested algorithms using the same dataset described in \cite{Suciu:2015} which encompasses data from February 2014 to January 2015. Significance was calculated using a 10-fold cross-validated paired t-test between the algorithm's and the baseline's f-score. P-values greater than 0.05 implies no significant improvement. We also used the f-score because, in our tests, the SVM was the only algorithm which had recall greater than the precision, so using either of these measures could lead to incorrect conclusions. The results reveal that, regardless of the algorithm, there is still room for improvement. We believe this can be achieved with modifications to the preprocessing and the data-gathering technique. We'll discuss that in Section \ref{sec:conclusion-future}. Next, we analyze each of the used algorithms.

\begin{table}[h!]
\scalefont{.65}
    \centering
    \begin{tabular}{|r|c|c|c|c|} \hline
        &					\textbf{Precision} & 	\textbf{Recall} & 	\textbf{F-score} & \textbf{P-value} \\ \hline
        Baseline - SVM (PoC)	&		0.2075	&	0.7053 &	0.3199 & --- \\ \hline
        LR (PoC)            &			0.6678	&	0.2434 &	0.3568 & 0.252\\ \hline
        XGBoost (PoC)       &			0.7454	&	0.2746 &	0.4014 & 0.078\\ \hline
        LightGBM (PoC)      &			0.7170	&	0.3293 &	0,4513 & $<$ 0.001\\ \hline\hline
        
        Baseline - SVM(RW)  &		    0.0632	&	0.7660 &	0.1166 & --- \\ \hline
        LR (RW)	            &			0.7		&	0.1857 &	0.2935 & 0,007\\ \hline
        XGBoost (RW)	    &			0.4916	&	0.1535 &	0.2340 & 0,080\\ \hline
        LightGBM (RW)	    &			0.5219	&	0.2196 &	0.3091 & 0,004\\ \hline

    \end{tabular}
    \textbf{\caption{\label{table:alg-results} Overall results}}
\end{table}

\subsubsection{SVM (Baseline)}

\begin{figure*}[!h] 
  \centering
  \begin{subfigure}[b]{0.32\linewidth}
    \includegraphics[width=\linewidth]{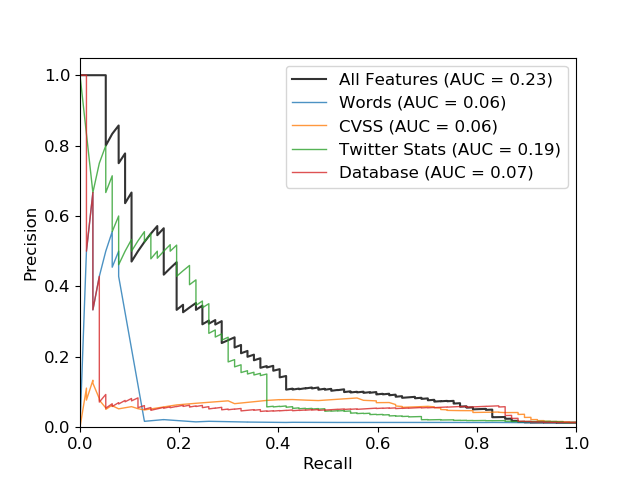}
    \caption{SVM - real-world exploits}
  \end{subfigure}
  \begin{subfigure}[b]{0.32\linewidth}
    \includegraphics[width=\linewidth]{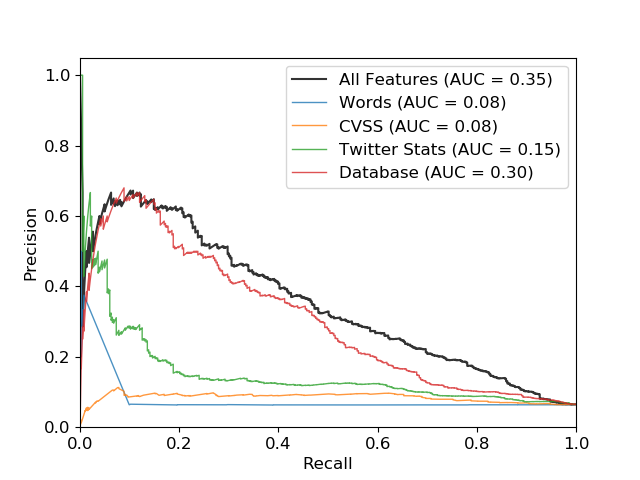}
    \caption{SVM - PoC exploits}
  \end{subfigure}
    \begin{subfigure}[b]{0.32\linewidth}
    \includegraphics[width=\linewidth]{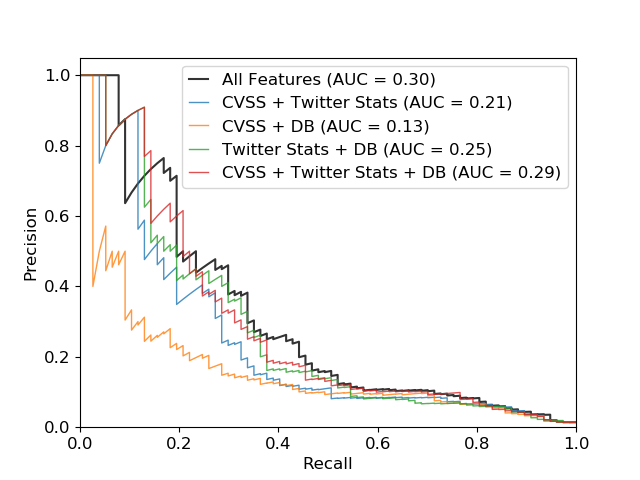}
    \caption{LG - real-world exploits}
  \end{subfigure}
  
  \caption{\label{fig:svm}Precision-recall curves for SVM and Logistic Regression}
\end{figure*}

Fig. \ref{fig:svm} shows the difference between the Precision-recall curves for PoC and real-world exploits. The first characteristic we would like to highlight is how the classifier performs better with PoC data, probably because vulnerabilities with exploits published on the ExploitDB are likely to have references on their description on NVD or OSDB. Therefore the subset of features extracted from public vulnerabilities databases plays a big part in the overall performance for PoC (this holds for all other algorithms). 

Secondly, it is possible to notice how, in real-world scenarios, the Twitter Statistics subset has an essential contribution to the overall result. As will be demonstrated here, this characteristic depends on the algorithm. However, our results indicate that the ``who said it'' might be a more relevant question than ``what was said'' to determine if a tweet indicates an exploited vulnerability.

Another characteristic is how the CVSS tends to be a conservative indicator. In other words, using just that score leads to labeling most of the vulnerability as \emph{exploited}, hence the low precision and high recall. On the other hand, Words subset tends to be the other way around: higher precision but very low recall. This behavior reflects how certain words on the BoW representation, such as  ``exploit'' or ``beware'', are particularly efficient in detecting exploits but are generally related to only a subset of the exploits.

\subsubsection{Logistic Regression (LG)}

Considering the precision, the Logistic Regression (LG) was the best performing algorithm for the RW scenario. When looking at F-score, the Logistic Regression was the second best. Fig. \ref{fig:svm} c) shows the PR curves for our test. In it, we display different combinations of features. We can highlight how Twitter statistics was useful for all the tests it was included, once again. It is worth mentioning that the model was one of the fastest to train and test.

\subsubsection{XGBoost and LightGBM}

\begin{figure*}[hb!]
  \centering
  \begin{subfigure}[b]{0.31\linewidth}
    \includegraphics[width=\linewidth]{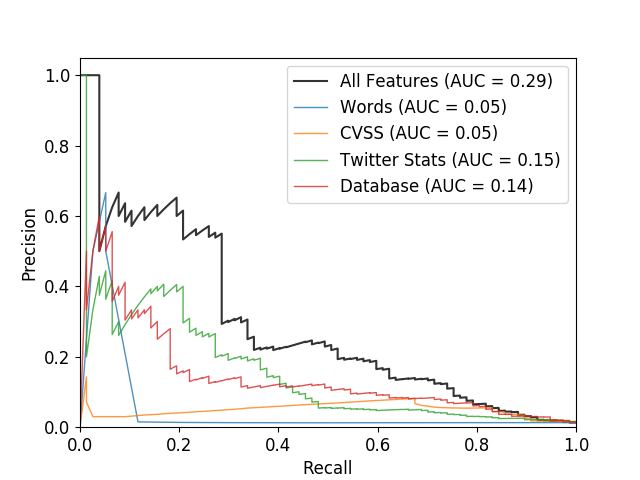}
    \caption{PR Curves for the XGBoost with RW exploits}
  \end{subfigure}
  \begin{subfigure}[b]{0.31\linewidth}
    \includegraphics[width=\linewidth]{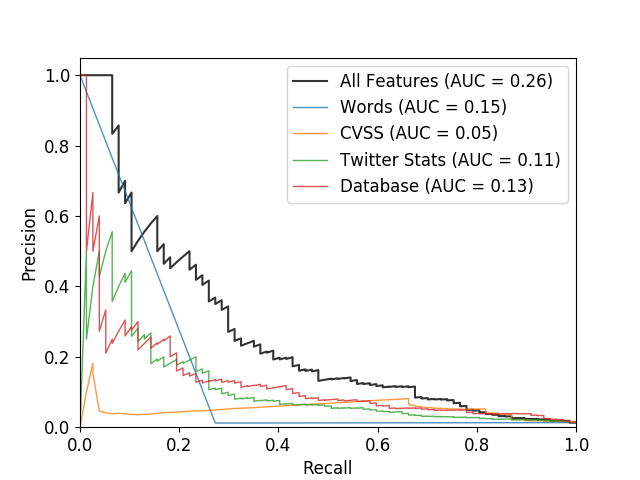}
    \caption{PR Curves for the LightGBM with RW exploits}
  \end{subfigure}
  \begin{subfigure}[b]{0.31\linewidth}
    \includegraphics[width=\linewidth]{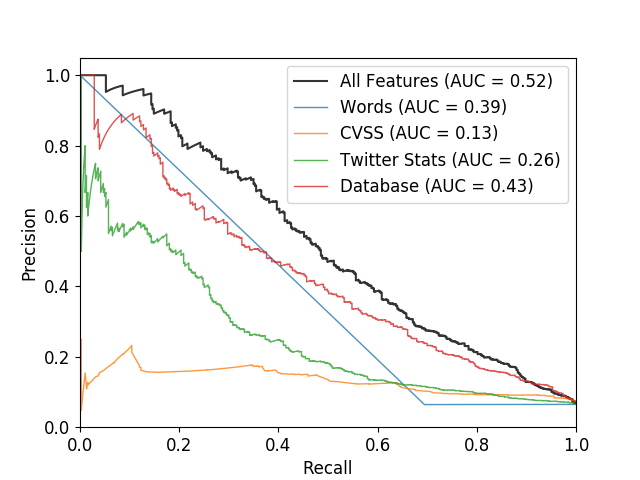}
    \caption{PR Curves for the LightGBM with PoC exploits}
  \end{subfigure}
  
  \caption{\label{fig:xgboost-lightgbm}Precision-recall curves for XGBoost and LightGBM}
\end{figure*}

Both of the ensembles tested had good results, but only one of them, the LightGBM, had a P-value of less than 0.05. Besides, the XGBoost demonstrated the second-longest execution time, ahead of only the SVM. The LightGBM, on the other hand, was the fastest algorithm to execute in our tests. Figure \ref{fig:xgboost-lightgbm} shows the PR curves for both algorithms.

In conclusion, the LightGBM was the best performing algorithm, considering the F-score, during our initial tests by a small margin, followed by the Logistic Regression. Throughout our other tests, the LightGBM was able to improve more than the Logistic Regression, as will be shown in Section \ref{sec:res_class_balancing}. Finally, we would like to point out how the CVSS subset of features plays a more significant part when testing with PoC exploits. For RW exploits, the Twitter statistics and the NVD data become more relevant.

\subsection{Multiple Ground Truth}\label{sec:res_gts}

\begin{table}[h!]
\scalefont{.65}
    \centering
    \begin{tabular}{|l|r|r|r|r|r|g|}
    \hline
      & \multicolumn{1}{l|}{\textbf{2014}} & \multicolumn{1}{c|}{\textbf{2015}} & \multicolumn{1}{c|}{\textbf{2016}} & \multicolumn{1}{c|}{\textbf{2017}} & \multicolumn{1}{c|}{\textbf{2018}} & \multicolumn{1}{G|}{\textbf{Total}} \\ \hline
    \textbf{Symantec Tweeted} & \textbf{77*}        & \textbf{43}           & \textbf{29}            & \textbf{58}          & \textbf{131}      & \textbf{267}                \\ 
    \textit{- Symantec Total} & \textit{90*}        & \textit{261}          & \textit{247}           & \textit{219}         & \textit{364}      & \textit{\textbf{1.228}}     \\ \hline
    \textbf{Avast Tweeted}    & \textbf{112}        & \textbf{33}           & \textbf{3}             & \textbf{4}           & \textbf{4}        & \textbf{44}                 \\
    \textit{- Avast Total}    & \textit{123}        & \textit{220}          & \textit{97}            & \textit{21}          & \textit{8}        & \textit{\textbf{346}}       \\ \hline
    \textbf{Other Tweeted}    & \textbf{11}         & \textbf{5}            & \textbf{2}             & \textbf{14}          & \textbf{7}        & \textbf{28}                 \\
    \textit{- Others Total}   & \textit{14}         & \textit{19}           & \textit{3}             & \textit{15}          & \textit{7}        & \textit{\textbf{45}}        \\ \hline
    \textbf{PoC Tweeted}      & \textbf{383}        & \textbf{115}          & \textbf{90}            & \textbf{220}         & \textbf{257}      & \textbf{699}                \\
    \textit{- PoC Total}      & \textit{823}        & \textit{721}          & \textit{549}           & \textit{1124}        & \textit{981}      & \textit{\textbf{3.984}}     \\ \hline
    \multicolumn{7}{l}{*\cite{Suciu:2015} dataset} \\
    \end{tabular}
    \caption{\label{tab:results-multi-gt-avnuns} Number of exploited vulnerabilities mentioned by vendors}
\end{table}

\begin{table*}[ht!]
\scalefont{.65}
    \centering
    \begin{tabular}{|r|c|c|c||c|c|c||c|c|c|}
    
     \multicolumn{10}{c}{  } \\
    \hline
								& \multicolumn{3}{c||}{SVM}         & \multicolumn{3}{c||}{LR}          & \multicolumn{3}{c|}{LightGBM}     \\ \hline
								& Precision & Recall & F-score      & Precision & Recall 	& F-score      	& Precision & Recall	& F-score \\ \hline
		Baseline               	& 0.2244  	& 0.8278 & 0.3531  		& 0.5850   	& 0.1094   	& 0.1844   		& 0.5458   	& 0.2844   	& \multicolumn{1}{G|}{0.3740}  \\ \hline
		ADASYN$^{\mathrm{I}}$  	& 0.2271  	& 0.8285 & 0.3565  		& 0.1936   	& 0.8403   	& 0.3147   		& 0.4640   	& 0.3774   	& 0.4162  \\ \hline
		SMOTE$^{\mathrm{I}}$   	& 0.2391  	& 0.8044 & 0.3686  		& 0.2200   	& 0.8181   	& 0.3467   		& 0.4867   	& 0.3785   	& 0.4258  \\ \hline
		AllKNN$^{\mathrm{II}}$ 	& 0.2271  	& 0.8285 & 0.3565  		& 0.4955   	& 0.3007   	& 0.3743   		& 0.4956   	& 0.5267   	& \multicolumn{1}{G|}{\textbf{0.5107}}  \\ \hline
		RUS$^{\mathrm{II}}$    	& 0.2330  	& 0.8303 & 0.3639  		& 0.1322   	& 0.8921   	& 0.2302   		& 0.2212   	& 0.8642   	& 0.3523  \\ \hline

    \multicolumn{10}{l}{$^{\mathrm{I}}$Over-sampling technique} \\
    \multicolumn{10}{l}{$^{\mathrm{II}}$Under-sampling technique} \\
    
    \end{tabular}
    \caption{\label{tab:results-balancing} Results for class balancing for RW exploits}
\end{table*}

In this experiment, we investigate how using different antivirus signatures as ground truth interferes with the classifier efficiency. In our research, we were able to find public databases with lists and descriptions of signatures from the following vendors: Avast, ESET, Symantec, and Trend Micro. We then developed web crawlers to collect this information, searched for CVEs mentions, and created an expanded ground truth that will be shared with the general public.

To our knowledge, all previous works use only Symantec to tell if a vulnerability has been exploited in the wild. Our findings indicate that, from 2015 to 2018, at least 248 of the 1,338 real-world exploited vulnerabilities are not mentioned by Symantec. Furthermore, considering years before 2015, Symantec's database misses 518 real-world exploited vulnerabilities in a total of 2,655 mentioned in signatures descriptions. Notice that this does not necessarily mean their antivirus has no signature for those exploits, but it indicates that no reference was made in the description. Table \ref{tab:results-multi-gt-avnuns} shows the amount of exploited vulnerabilities mentioned by each vendor from 2015 to 2018. ESET and Trend Micro contain limited information compared to Avast and Symantec, so we combined their numbers and labeled them ``Other''. Also, notice how the quantities change drastically over time, we will discuss that at the end of this section.

\begin{figure}[h!]
    \centering
    \includegraphics[width=.35\textwidth]{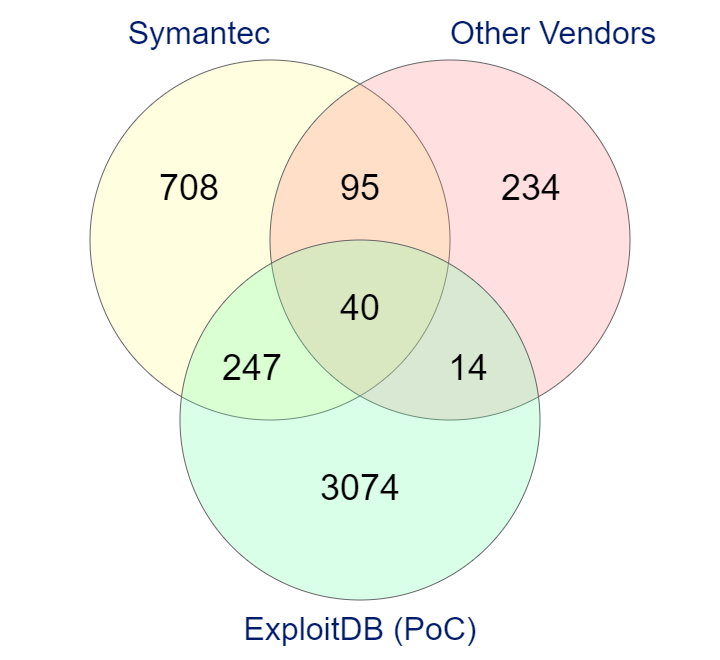}
    \caption{The intersection of the lists of exploited vulnerabilities}
    \label{fig:results-multi-gt-venn}
\end{figure}

One of our concerns about using public signature descriptions was to avoid vendors that would only include PoC information and count it as ``exploit detection''. To verify if we were getting new information, for each vendor, we compared the lists of exploited vulnerabilities to our list of PoC. We found that a significant amount of exploited vulnerabilities were nor mentioned by Symantec or EDB, so it is reasonable to consider them as real-world exploits. Figure \ref{fig:results-multi-gt-venn} shows the intersection of our new list of exploited vulnerabilities with Symantec's and EDB's (from 2015 to 2018).

To check how the new ground truth affects our classifier, we conducted a series of tests where the model was trained using labels from a single vendor but tested with the combination of all vendors (the combined ground truth). Table \ref{table:results-multi-gt} contains the precision, recall, and F-score for each test with our best performing algorithm, the LightGBM.

\begin{table}[h!]
\scalefont{.65}
    \centering
    \begin{tabular}{|l|c|c|c|c|}
    \hline
    \textbf{}                     & \textbf{Precision} & \textbf{Recall} & \textbf{F-score} & \textbf{P-value} \\ \hline
    \textbf{Symantec}             & 0.5605       & 0.1233    & 0.2021     & --        \\ \hline
    \textbf{Avast}                & 0.5242       & 0.2209    & 0.3109     & 0.086     \\ \hline
    \textbf{Sym + Avast}          & 0.5238       & 0.2795    & 0.3645     & 0.046     \\ \hline
    \textbf{Sym + Avast + Others} & 0.5458       & 0.2844    & 0.3740     & 0.032     \\ \hline
    \end{tabular}
    \textbf{\caption{\label{table:results-multi-gt} Results with combined ground truth}}
\end{table}

The results suggest that in 2014, \cite{Suciu:2015} could have achieved equal or better results using Avast's database instead of Symantec's (the P-value larger than 0.05 means the improvement is not significant enough for us to declare it better, despite the F-score). Most importantly, we conclude that using information from a single vendor can bias the model toward some vulnerabilities and lead to non-optimal performance on real-world scenarios. As we will explain in Section \ref{sec:res_class_balancing}, the gains of a combined ground truth can even be emphasized when using a class balancing algorithm. Unfortunately, as shown in Table \ref{tab:results-multi-gt-avnuns}, since 2015, Avast, has gradually decreased the volume of information about its signatures.

From this experiment, we conclude that using multiple ground truth sources is crucial for any machine learning application dealing with exploit detectors. Despite that, we see less of this information available in recent years, which may indicate less effective classifiers on future works.

\subsection{Class-balancing}\label{sec:res_class_balancing}

In another experiment, we compared methods for dealing with class imbalance. The severe imbalance in our application motivated this test, as shown in Table \ref{table:qntd-cves}. In this test, we ran class balancing methods with all four classification algorithms, and once again, the LighGBM outperformed the others. We used a Python library called ``imbalanced-learn'' and tested several algorithms of both undersampling and oversampling. In our tests, we used the combined ground truth from Section \ref{sec:res_gts} on real-world exploits and executed these algorithms for the training set from each fold of the 10-fold cross-validation. Table \ref{tab:results-balancing} shows the results for the best-performing ones.

As we can see, the SVM and the LightGBM behave differently with the balancing algorithms. While the baseline could only get similar F-score values and no statistically significant improvement, the LightGBM showed a considerable increase in performance, with a P-value of 0.001, when used with the AllKNN, an under-sampling technique. Other algorithms were not able to obtain statistically different results, although some achieved superior F-score.

In general, the AllKNN outperformed the original experiments for all classification algorithms, both for PoC and real-world exploits. It is worth mentioning that when using only Symantec's ground truth, improvements were more modest, highlighting the importance of using multiple sources for ground truth.

\subsection{Updated Data and Time Window Sizes}\label{sec:res_time_window}

With our last test, we wanted two things. First, to understand if the results obtained using data from 2014 would remain with an updated dataset. Second, to answer if training with data from a more extended period would influence the results. To do that, we first trained and tested the model separating our data by year, from 2015 to 2018. We then ran a series of experiments where the model trained with different time windows from years before 2018 but tested with data from 2018. More specifically, the model trained with the following groups of years: 2017, then from 2016 to 2017, and finally, from 2015 to 2017. Here, we would like to evaluate whether adding more training data would positive influence the performance of the model.In this experiment, we used the LightGBM with the combined ground truth. The results are shown in Table \ref{tab:results-year-by-year} and Table \ref{tab:results-time-window}.

\begin{table}[h!]
\scalefont{.8}
    \centering
    \begin{tabular}{|r|c|c|c|}
    \hline
    Year & Precision & Recall & F-score \\ \hline
    2015 & 0.6048    & 0.6282 & 0.6163  \\ \hline
    2016 & 0.2089    & 0.2333 & 0.2204  \\ \hline
    2017 & 0.5419    & 0.5681 & 0.5547  \\ \hline
    2018 & 0.6999    & 0.5529 & 0.6178  \\ \hline
    \end{tabular}
    \caption{\label{tab:results-year-by-year} Results for updated data - Training and testing using a one-year window}
\end{table}

\begin{figure}[h!]
    \centering
    \includegraphics[width=.40\textwidth]{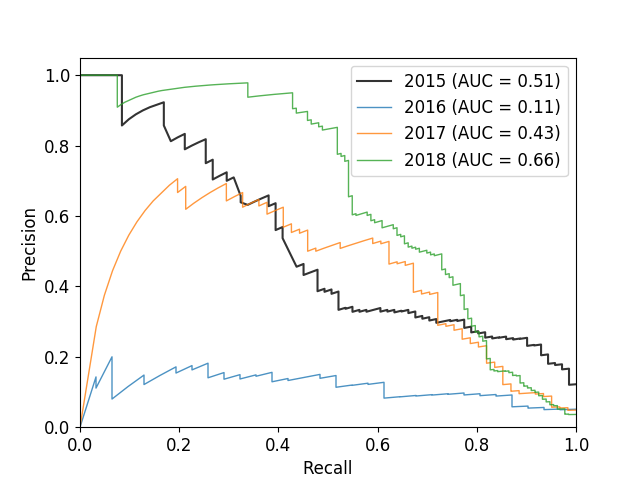}
    \caption{Precision-recall curves for each year on RW exploits (2015 - 2018)}
    \label{fig:results-pr-by-year}
\end{figure}

\begin{table}[h!]
\scalefont{.65}
    \centering
    \begin{tabular}{|l|c|c|c|c|}
    \hline
    \multicolumn{1}{|c|}{Train}           & Test & Precision & Recall & F-score \\ \hline
    2018 (baseline) & 2018 & 0.6999    & 0.5529 & 0.6178                       \\ \hline
    2017            & 2018 & 0.3333    & 0.1654 & 0.2211                       \\ \hline
    2016 2017       & 2018 & 0.3051    & 0.1353 & 0.1875                       \\ \hline
    2015 2016 2017  & 2018 & 0.3208    & 0.1278 & 0.1828                       \\ \hline
    \end{tabular}
    \caption{\label{tab:results-time-window} Results for training with different time windows}
\end{table}

We can see from Table \ref{tab:results-time-window} that results remain reasonably constant throughout the years, except for 2016, probably due to the small number of exploited vulnerabilities mentioned on Twitter (see Table \ref{tab:results-multi-gt-avnuns}). Moreover, as mentioned in Section \ref{sec:res_gts}, the contribution of Avast and other antiviruses to our ground truth diminished in recent years, making the model more dependent on Symantec as its only source of information. 

When training with different time windows, our results suggest that there may be some relation between CVEs from the same year that allows better learning from the model. When training with data from past years, F-score remained around 0.2. From our perspective, this difference was not expected since all data used for the model's features were equally available for all of the years. To check if information learned by the model was getting old overtime or if it was missing new information, we ran tests with multiple combinations of time windows (e.g., we would train just with 2015 or with 2015 and 2016, then with 2016 and 2017; we also tested with multiple years, and even trained with years ahead of the testing one). In all cases, the performance was significantly lower than the single year approach. We believe this behavior can relate to the fact that some vulnerabilities are disclosed together and be part of the same malware issue. The WannaCry ransomware outbreak, for example, was related to six different CVEs (CVE-2017-0143, CVE-2017-0144, CVE-2017-0145, CVE-2017-0146, CVE-2017-0147, and CVE-2017-0148), all with a fairly similar description, CVSS, and affecting the same products. If that is the case, maybe it is necessary to detect these groups of CVEs and treat them as a single vulnerability. To conclude, we found no empirical evidence that using more than one year of data can benefit a model that uses a temporal batch split.

\subsection{Threats to validity}\label{sec:threats-to-validity}

Our experiments were conducted with tweets acquired using a tool called GetOldTweets3\cite{getoldtweets}, which can search for Twitter messages regardless of their posting time but cannot obtain deleted tweets or tweets from deleted users. In contrast, if we were using the Twitter Stream API, messages would be obtained and stored in our database when posted. This limitation leads to a diminished tweet volume. We believe our contributions can apply for batch training with or without a temporal split, but it is necessary to mention that time intermixing can be a limitation on our tests using single year data. In our method, we also discarded tweets referencing older vulnerabilities (for example, we only consider CVEs disclosed in 2018 when collecting Twitter mentions in 2018), which may prevent the system from detecting new exploits on old vulnerabilities.

\section{Conclusion and Future Work}\label{sec:conclusion-future}

In this paper, we explored several aspects of Twitter-based exploit detectors. We have compared the performance of four commonly used classification algorithms, conducted tests with different data sources for ground truth to real-world exploits, applied balancing algorithms, and trained and tested our classifier with different time windows.



By selecting a suitable algorithm, managing class imbalance issues, and adding new ground truth, we were able to outperform the baseline and evaluate the performance of exploit detectors using updated data. Some of our results raised questions that we would like to tackle in future work, some of them involve: monitoring the NVD data to check which of our features is available when a CVE is published; understanding how related CVEs interfere on our classifier and if that is the real cause of inferior performance using time windows; enhance the tweet searching by using keywords other than ``CVE''. We also want to experiment with different feature selection and text representations.


\bibliographystyle{sbc}
\bibliography{sbc-template}

\end{document}